\documentclass[10pt,letterpaper,compsoc,conference]{iiswc25}

\usepackage{cite}
\usepackage{amsmath,amssymb,amsfonts}
\usepackage{graphicx}
\usepackage[dvipsnames]{xcolor}
\usepackage[final]{microtype}
\usepackage[italic]{mathastext}
\usepackage{libertine}
\usepackage[T1]{fontenc}
\usepackage{textcomp}
\usepackage[varqu,varl]{zi4}
\usepackage[all]{nowidow}
\usepackage[keeplastbox]{flushend}
\usepackage{fancyhdr}

\usepackage{listings}
\usepackage{xcolor}

\definecolor{codegreen}{rgb}{0,0.6,0}
\definecolor{codegray}{rgb}{0.5,0.5,0.5}
\definecolor{codepurple}{rgb}{0.58,0,0.82}
\definecolor{backcolour}{rgb}{0.95,0.95,0.92}

\lstdefinestyle{mystyle}{
    backgroundcolor=\color{backcolour},   
    commentstyle=\color{codegreen},
    keywordstyle=\color{magenta},
    numberstyle=\tiny\color{codegray},
    stringstyle=\color{codepurple},
    basicstyle=\ttfamily\footnotesize,
    breakatwhitespace=false,         
    breaklines=true,                 
    captionpos=b,                    
    keepspaces=true,                 
    numbers=left,                    
    numbersep=5pt,                  
    showspaces=false,                
    showstringspaces=false,
    showtabs=false,                  
    tabsize=2
}

\usepackage{tcolorbox}
\usepackage{subfig}
\usepackage{array}
\newcolumntype{L}{>{\centering\arraybackslash}m{0.4in}}
\usepackage{caption}
\usepackage{wrapfig}
\usepackage{algpseudocode}
\algrenewcommand{\algorithmiccomment}[1]{\hfill\textbf{//}\,#1}
\usepackage{algorithm}
\algrenewcommand\alglinenumber[1]{\small #1:}
\usepackage{circledsteps}
\pgfkeys{/csteps/inner color=white}
\pgfkeys{/csteps/fill color=black}
\usepackage{pifont}
\usepackage{multirow}
\usepackage{mdframed}

\usepackage{soul}

\edef\oldtt{\ttdefault}
\usepackage[scaled]{beramono}
\usepackage[T1]{fontenc}
\renewcommand{\ttdefault}{\oldtt}

\usepackage{mathtools}
\usepackage{threeparttable}

\fancypagestyle{firstpage}{
  \fancyhf{}
  
}

\begin{document}

%% EDIT TITLE BELOW

\title{
\vspace{-0.5em}
\textit{No One-Size-Fits-All}: A Workload-Driven Characterization of Bit-Parallel vs. Bit-Serial Data Layouts for Processing-using-Memory
}

%% DO NOT EDIT THE FOLLOWING

% \renewcommand\Authsep{\qquad}
% \renewcommand\Authand{\qquad}
% \renewcommand\Authands{\qquad}

%% EDIT AUTHOR LIST BELOW

\author{Jingyao Zhang, Elaheh Sadredini\\
\textit{Department of Computer Science and Engineering}\\
\textit{University of California, Riverside}}

% \author{Elaheh Sadredini}
%\author{Author3 Name}
% \affil{University of California,Riverside}

%%% ALTERNATIVE FORMAT FOR MULTIPLE SCHOOLS:
%%% 
% \author[1]{Author1 Name}
% \author[2]{Author2 Name}
% \author[2]{Author3 Name}
% \author[1]{Author4 Name}
% \affil[1]{Full Name of Awesome School}
% \affil[2]{Full Name of Awesomer School}

\maketitle
\thispagestyle{firstpage}
\pagestyle{plain}

%% EDIT YOUR PAPER'S CONTENTS BELOW

\begin{abstract}
Processing-in-Memory (PIM) is a promising approach to overcoming the memory-wall bottleneck. However, the PIM community has largely treated its two fundamental data layouts, Bit-Parallel (BP) and Bit-Serial (BS), as if they were interchangeable. This implicit "one-layout-fits-all" assumption, often hard-coded into existing evaluation frameworks, creates a critical gap: architects lack systematic, workload-driven guidelines for choosing the optimal data layout for their target applications.
    To address this gap, this paper presents the first systematic, workload-driven characterization of BP and BS PIM architectures. We develop iso-area, cycle-accurate BP and BS PIM architectural models and conduct a comprehensive evaluation using a diverse set of benchmarks. Our suite includes both fine-grained microworkloads from MIMDRAM to isolate specific operational characteristics, and large-scale applications from the PIMBench suite, such as the VGG network, to represent realistic end-to-end workloads.
    Our results quantitatively demonstrate that no single layout is universally superior; the optimal choice is strongly dependent on workload characteristics. BP excels on control-flow-intensive tasks with irregular memory access patterns, whereas BS shows substantial advantages in massively parallel, low-precision (e.g., INT4/INT8) computations common in AI. Based on this characterization, we distill a set of actionable design guidelines for architects. This work challenges the prevailing one-size-fits-all view on PIM data layouts and provides a principled foundation for designing next-generation, workload-aware, and potentially hybrid PIM systems.
\end{abstract}

\section{Introduction}

The growing mismatch between processor speed and memory bandwidth, commonly referred to as the \textit{memory wall}, has become one of the most pressing challenges in modern computing. As applications scale in data intensity, the cost of moving data between memory and computation units increasingly dominates overall system performance and energy consumption. Recent studies show that data movement can account for more than 60\% of system energy in large-scale applications~\cite{Boroumand2018-ue}. Processing-in-Memory (PIM) has emerged as a promising architectural approach to mitigate this challenge by relocating computation closer to where data resides.

PIM architectures can be broadly divided into two categories: Processing-Near-Memory (PNM) and Processing-Using-Memory (PUM). PNM places general-purpose or specialized compute logic near the memory periphery, either within memory modules, in the controller, or on the logic die in 3D-stacked DRAM, as demonstrated in academia (Chameleon~\cite{AsghariMoghaddam2019-chameleon}, RecNMP~\cite{Liu2021-recNMP}, and Tesseract~\cite{Oliveira2017-ARC}) and commercial systems (Samsung's HBM-PIM~\cite{Lee2021-lu} and UPMEM~\cite{Devaux2019TheTP}). These systems benefit from reduced memory access latency and high bandwidth but still suffer from limited internal memory parallelism. In contrast, PUM architectures exploit the physical properties of memory cells to compute \textit{in situ}, directly within the memory arrays. By turning memory into a compute substrate, PUM systems can eliminate processor-memory communication overhead and enable massive parallelism and energy efficiency. DRAM-based PUM systems such as Ambit~\cite{Seshadri2017-ej}, DRISA~\cite{Li2017-nn}, pLUTo~\cite{Ferreira2021-pLUTo}, and PULSAR~\cite{Yuksel2023-PULSAR} demonstrate how bulk bitwise operations and table lookups can be efficiently executed in commodity DRAM. Similarly, SRAM-based designs like Neural Cache~\cite{Eckert2018-cl} and Compute Caches~\cite{Aga2017-kx} enable reconfigurable or bit-serial logic execution directly in SRAM arrays. Non-volatile memory (NVM) technologies, such as ReRAM and PCM, also support compute-in-memory through techniques like PRIME~\cite{Chi2016-ISCA} and Pinatubo~\cite{Li2016-DAC}, which exploit resistive switching to perform parallel logic operations.

While PUM architectures have made significant progress, one critical but underexplored design choice in PUM remains the organization of data within memory arrays. At the heart of every PUM system is a fundamental decision: whether to adopt a \textbf{Bit-Parallel (BP)} layout, in which multi-bit words are stored horizontally across multiple bitlines to support word-level parallelism; or a \textbf{Bit-Serial (BS)} layout, where bits of a word are stored vertically along a single column, enabling massive bit-level parallelism. This choice has profound implications for performance and resource utilization.

Many works adopt layout strategies driven by specific hardware constraints or workload assumptions, without systematic justification. For example, BitFusion~\cite{Sharma2018-qm} adopts a BS layout to exploit fine-grained bit-level reuse for DNN inference, while PRIME~\cite{Chi2016-vn} uses a BP layout to enable analog matrix-vector multiplication in ReRAM crossbars. Ambit~\cite{Seshadri2017-ej} supports both layouts depending on the type of operation but offers no framework for guiding layout selection. Other architectures like BP-NTT~\cite{Zhang2023-rt}, DRISA~\cite{Li2017-nn}, Cascade~\cite{Chou2019-zh}, and FloatPIM~\cite{Imani2019-zv} also make fixed layout choices without articulating how these decisions affect broader system behavior across workloads. In many cases, these decisions are optimized for a narrow class of operations or tied to hardware capabilities, with little discussion of trade-offs or alternatives. Even comprehensive surveys~\cite{Mutlu2019-uv, Singh2018-dt} that aim to organize the PIM design space treat data layout as a backend implementation detail, overlooking its first-order effects on programmability, control overhead, data reuse, and hardware efficiency. 

Benchmarking tools further reinforce this oversight: for instance, PIMeval~\cite{Siddique2024-ip}, a state-of-the-art benchmarking framework for PIM, statically associates data layout with hardware templates; for example, selecting a BitSIMD-V device enforces the use of a bit-serial (vertical) layout for all operations, irrespective of the specific workload characteristics. This rigid assumption fails to reflect the diverse computational needs of real-world applications and conceals potential inefficiencies introduced by mismatched layout-workload pairings. To illustrate the impact of this assumption, we conduct an evaluation of VGG-13 inference on a 512-column SRAM-based PIM array configured with a bit-serial layout. In the fully connected layers, where only 8 output neurons are active, the limited degree of parallelism results in severe underutilization—only 5.5\% of the available compute columns are used. This inefficiency leads to wasted cycles and elevated energy per operation, underscoring the need for layout-aware workload mapping.

%Moreover, layout-related limitations are exacerbated when supporting mixed-precision or control-intensive workloads. For instance, BS layouts struggle with managing multiple bit slices for 16-bit or 32-bit arithmetic, requiring repeated synchronization, carry propagation, and transposition across columns. Conversely, BP layouts often allocate unnecessary storage and switching energy for unused upper bits in quantized or sparse computation. Without explicit design-time evaluation, these overheads remain hidden until deployment.

In summary, while existing PUM systems implement a range of data layout strategies, they largely lack systematic analysis or data-driven guidance for layout selection. This has fostered a ``one-size-fits-all'' mindset, where layout choices are fixed without considering workload-specific requirements, leading to inefficiencies and suboptimal designs. This implicit assumption reveals itself in three key limitations. First, despite numerous proposals, no systematic study has characterized how BP and BS layouts differ in their fundamental capabilities. For instance, while bit-serial excels at massively parallel bit-wise operations common in binary neural networks~\cite{Rastegari2016-ym}, it suffers from severe row overflow when storing multiple word-level intermediate results, a common pattern in iterative algorithms like FIR (Finite Impulse Response) filters. Conversely, bit-parallel naturally handles such scenarios with superior efficiency. However, when workloads exhibit massive bit-level parallelism that exceeds the array's PE count, such as computing hamming distances across thousands of binary vectors, bit-serial's ability to use every column as an independent 1-bit ALU provides a fundamental advantage. Second, existing PIM evaluations typically focus on narrow benchmark suites that favor one layout over another. Studies using only CNN workloads~\cite{Chen2016-le,Kwon2018-ml} miss BP's advantages in control-flow-intensive code, while those focusing on traditional CPU benchmarks~\cite{Ahn2016-wc} overlook BS's efficiency for low-precision AI inference. The recent PIMBench suite~\cite{Siddique2024-ip} provides diverse workloads but lacks layout-aware analysis. This fragmented evaluation landscape prevents architects from understanding the true performance trade-offs. Third, architects lack principled heuristics for layout choice. Questions like "should a PIM accelerator for edge AI use bit-serial?" or "When does the overhead of supporting both layouts, as in RecNMP~\cite{Ke2020-xs}, pay off?" remain unanswered, leading to ad-hoc, suboptimal design choices.

This paper presents the first systematic, workload-driven characterization of bit-parallel versus bit-serial data layouts in PIM architectures. Rather than advocating for one layout over another, we demonstrate that \textbf{optimal layout selection is fundamentally workload-dependent}—and that ignoring this dependency incurs substantial performance penalties.
Our key insight is that different computational patterns exhibit natural affinities to different layouts. 
We identify six fundamental challenges inherent to the BS data layout that lead to significant inefficiencies in common computational scenarios and demonstrate how adopting a Bit-Parallel (BP) data layout naturally and efficiently resolves them. We then distill these specific issues into four architectural root causes, providing a principled foundation for a more workload-driven approach to PIM design.%These architectural differences translate directly to workload performance. For example, a 4-tap FIR filter that elegantly fits in BP's row-wise storage would require complex multi-column coordination in BS due to row overflow.

To quantify these effects, we base our study on SRAM-based PIM as a representative substrate and develop cycle-accurate models of both BP and BS architectures under equal-area constraints, ensuring a fair comparison within identical silicon budgets. Unlike prior work that models only computation~\cite{Angizi2018-gq}, we explicitly account for bit-transposition overhead in BS designs, a cost often ignored but crucial for fair comparison~\cite{Hajinazar2021-tu}. Our evaluation employs a carefully curated two-tier benchmark suite: (1) microbenchmarks from MIMDRAM~\cite{Oliveira2024-fr} that isolate specific computational patterns, and (2) full applications from PIMBench~\cite{Siddique2024-ip} that represent real-world workloads spanning AI inference, graph processing, and data analytics.

This paper makes the following contributions: 

\begin{itemize}
\item \textbf{First Principled Characterization}: We provide the first systematic analysis of how bit-parallel and bit-serial data layouts in PIM architectures differ across multiple dimensions such as compute granularity, storage efficiency, control flow handling, and mixed-precision support.

\item \textbf{Quantitative Performance Analysis}: Through cycle-accurate modeling of iso-area BP and BS designs, we quantify performance differences across diverse workloads, revealing up to 14$\times$ variations between static layouts. Furthermore, hybrid execution that dynamically switches between BP and BS achieves up to 2.66$\times$ speedup over the best static choice.

\item \textbf{Workload-Aware Design Framework}: We develop a comprehensive taxonomy that maps workload characteristics (parallelism degree, data precision, control complexity) to optimal layout choices, synthesizing our findings into a classification framework that guides architects in selecting BP, BS, or hybrid approaches for their target applications.

\end{itemize}

The remainder of this paper is organized as follows. Section \ref{sec:background} provides background on PIM architectures and data layouts. Section \ref{sec:arch} presents our architectural analysis of BP versus BS designs. Section \ref{sec:methodology} describes our experimental methodology. Section \ref{sec:eval} presents our evaluation results across the benchmark suite. Section \ref{sec:conclusion} concludes.
\section{Background}
\label{sec:background}

Processing-in-Memory (PIM) has emerged as a compelling solution to the data movement bottleneck that characterizes modern computing. The PIM architectural landscape is broadly divided into two main categories: Processing-Near-Memory (PNM) and Processing-Using-Memory (PUM). PNM architectures integrate computational logic onto the memory controller or into the logic layer of 3D-stacked memory devices, as seen in commercial systems like Samsung's HBM-PIM~\cite{Lee2021-lu} and UPMEM~\cite{Gomez-Luna2021-mu}. While effective at reducing data movement to and from the host CPU, PNM still treats the memory array as a passive storage unit. In contrast, PUM architectures leverage the physical properties of the memory array itself to perform massively parallel, bit-level computations directly where data resides~\cite{Seshadri2017-ej, Chi2016-vn}. This approach offers the highest potential for parallelism and energy efficiency. Our work focuses on PUM, as the fundamental choice of data layout within the memory array---a choice with profound architectural implications---is most critical and impactful in this context.

\subsection{In-SRAM Computing Primitives}\label{sec:insram}
\begin{figure}[!t]
\centerline{\includegraphics[width=2.0in]{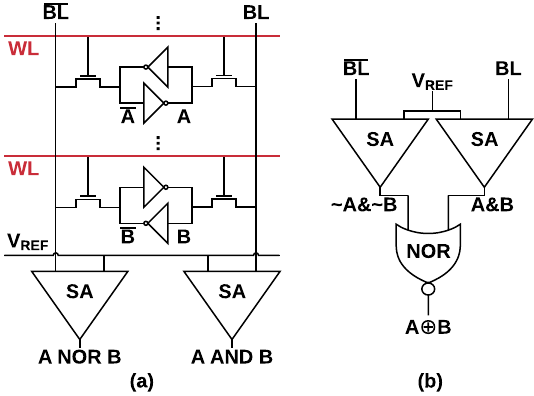}}
\caption{In-SRAM bitline operations. 
Simultaneous activation of two wordlines accomplishes AND/NOR operations, while an extra NOR gate facilitates XOR operation.
}
\label{fig:bitline}
\end{figure}

PUM architectures have been built on various memory technologies, including DRAM~\cite{Seshadri2017-ej} and NVM~\cite{Chi2016-vn}. For clarity, we use SRAM-based PUM as a representative example to illustrate the fundamental computational primitives. In these systems, logic operations are executed by exploiting the intrinsic analog behavior of the memory cell array. This technique, often called in-SRAM computing, capitalizes on activating multiple wordlines simultaneously instead of just one~\cite{Jeloka2016-na}. As shown in Figure~\ref{fig:bitline}, this causes the SRAM cells on the selected rows to jointly discharge the bitlines (BL and $\overline{\text{BL}}$). The resulting voltage, when measured by a sense amplifier (SA), corresponds to a bit-wise logical operation across the data stored in the activated rows.

For example, an element-wise AND is realized if the SA detects a high voltage on the BL, which only occurs if all participating cells store a `1'. A NOR operation is similarly realized on the complementary bitline ($\overline{\text{BL}}$). By combining these native AND/NOR capabilities, more complex operations like XOR can also be constructed, as shown in Figure~\ref{fig:bitline}(b). This massively parallel computation occurs simultaneously on all columns of the array, forming the basis of PUM's efficiency. Several research efforts have built upon these primitives. For instance, Compute Cache~\cite{Aga2017-kx} modifies the SA to support a richer set of logic operations, while Cache Automaton~\cite{Subramaniyan2017-fh} uses a sense-amplifier cycling mechanism to accelerate specific computations. Understanding these computational primitives is key to appreciating the profound impact of data layout, which we discuss next.

\subsection{Hierarchical Data Layouts}\label{sec:layouts}
\begin{figure}[t]
    \centering
    \includegraphics[width=3.4in]{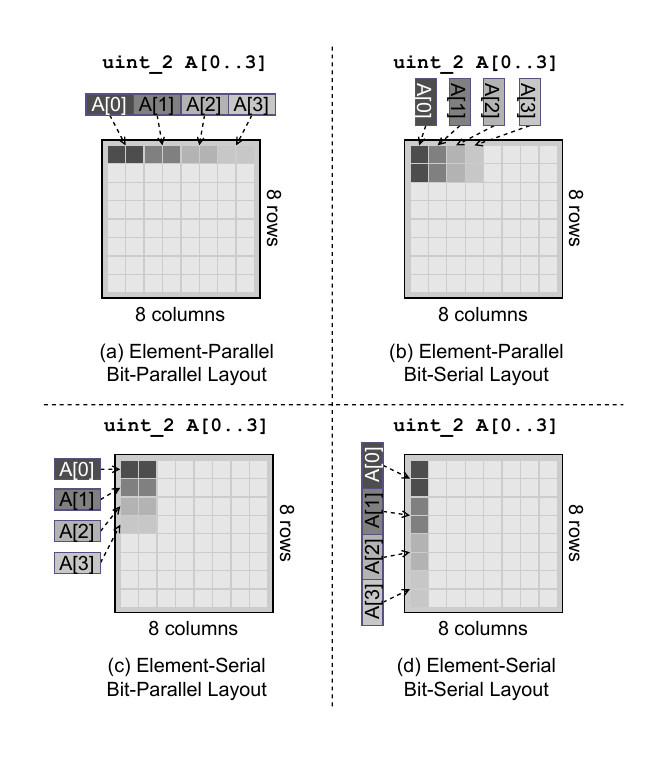} % Placeholder for the unified 4-part figure
    \caption{The four hierarchical data layout schemes resulting from combining bit-level (BP, BS) and vector-level (EP, ES) organization. (a) EP-BP: Ideal for inter-vector operations on wide data. (b) EP-BS: Maximizes parallelism for inter-vector operations. (c) ES-BP: Efficiently buffers a single vector for intra-vector operations. (d) ES-BS: Prone to row overflow for all but the simplest intra-vector tasks.}
    \label{fig:layout_combinations}
\end{figure}

At the core of PUM design, the data layout decision bifurcates into two primary strategies for organizing data words: \textbf{Bit-Parallel (BP)} and \textbf{Bit-Serial (BS)}. This choice dictates how bits of a single word are mapped to the 2D memory grid. Orthogonal to this is the vector-level organization, which determines how multiple data elements are arranged for processing---either in \textbf{Element-Parallel (EP)} or \textbf{Element-Serial (ES)}. The interplay of these choices yields four distinct layout strategies, shown in Figure~\ref{fig:layout_combinations}.

In \textbf{Bit-Parallel (BP)} layouts, an entire data word is stored horizontally across multiple columns (Figures~\ref{fig:layout_combinations}a and \ref{fig:layout_combinations}c). This design, which facilitates word-level operations, has been adopted by several PUM systems such as Compute Cache~\cite{Aga2017-kx}, Ambit~\cite{Seshadri2017-ej}, Sealer~\cite{Zhang2022-cm}, BP-NTT \cite{Zhang2023-rt} and Inhale~\cite{Zhang2022-vs}. The BP approach is effective for both traditional SIMD-style inter-vector operations (EP-BP) and for creating efficient intra-vector scratchpads (ES-BP).

In \textbf{Bit-Serial (BS)} layouts, an N-bit word is stored vertically down a single column (Figures~\ref{fig:layout_combinations}b and \ref{fig:layout_combinations}d). This approach has been the predominant strategy in PUM research, maximizing the number of parallel processing elements. It is the foundation for systems like Neural Cache~\cite{Eckert2018-cl}, Duality Cache~\cite{Fujiki2019-fi}, SIMDRAM~\cite{Hajinazar2021-tu}, MIMDRAM \cite{Oliveira2024-fr}, and Infinity Stream~\cite{Wang2023-az}. While highly effective for massively parallel inter-vector operations (EP-BS), it is often impractical for storing vectors of even moderate length (ES-BS) due to the limited physical depth of memory columns, a problem we term \textit{row overflow}.

\paragraph{Scope beyond SRAM.}  Although our analysis and models instantiate SRAM arrays, the core layout trade-offs (parallelism versus density, vertical storage pressure, lockstep control) are technology-agnostic and often intensify in non-volatile memories due to higher write latency and limited endurance. Therefore, the workload-driven conclusions and the benefits of hybrid layout support are expected to generalise to ReRAM/PCM PUM fabrics as well.

Despite the prevalence of the BS paradigm, recent works like Proteus~\cite{Oliveira2025-nw} and the PIMeval framework \cite{Siddique2024-ip} have demonstrated the growing relevance of BP for specific workloads, particularly those requiring dynamic precision or complex intra-vector operations. However, the fundamental trade-offs between BP and BS architectures have not been systematically analyzed. This paper provides the first direct, comprehensive comparison of these two competing PUM design philosophies, aiming to guide future architects in navigating this critical design choice.

\section{Architectural Principles: Challenges and Solutions}
\label{sec:arch}

The Bit-Serial (BS) paradigm is foundational to many influential PUM systems, including Neural Cache~\cite{Eckert2018-cl}, Duality Cache~\cite{Fujiki2019-fi}, SIMDRAM~\cite{Hajinazar2021-tu}, MIMDRAM~\cite{Oliveira2024-fr}, and Infinity Stream~\cite{Wang2023-az}, making it the de facto standard for achieving fine-grained parallelism. Despite its widespread adoption, our analysis reveals six fundamental challenges inherent to the BS data layout that lead to significant inefficiencies in common computational scenarios. This section systematically details these challenges and demonstrates how adopting a Bit-Parallel (BP) data layout naturally and efficiently resolves them. We then distill these specific issues into four architectural root causes, providing a principled foundation for a more nuanced, workload-driven approach to PIM design.

\subsection{Analysis Framework and Assumptions}
To provide a concrete and rigorous comparison, we ground our analysis in a set of common architectural parameters and cycle-accurate performance models for both BP and BS execution.

\textbf{Architectural Parameters.} We assume a conventional PIM system composed of SRAM arrays enabled for in-situ computing. The key parameters, representative of typical designs~\cite{Aga2017-kx, Seshadri2017-ej}, are a physical dimension of 128 rows (\texttt{array\_rows}) and 512 columns (\texttt{array\_columns}) per array, as detailed in Table~\ref{tab:arch_params}.

\begin{table}[h]
\centering
\caption{Architectural Parameters for Analysis.}
\label{tab:arch_params}
\scalebox{1.2}{
\begin{tabular}{l|c}
\hline
\textbf{Parameter} & \textbf{Value} \\ \hline
\texttt{array\_rows} & 128 \\
\texttt{array\_columns} & 512 \\ \hline
\end{tabular}
}
\end{table}

\textbf{Performance Models.} The total latency of a kernel is the sum of load, compute, and readout cycles. While our architectural analysis in this section focuses on the compute cycles to isolate fundamental operational differences, our complete model accounts for all overheads including bit-transposition costs for BS designs and data layout conversion overheads when switching between BP and BS representations—critical factors often overlooked in prior work. Table~\ref{tab:cycle_costs} defines the cycle costs for primitive compute operations. BP operates on word-level data, where logical operations (AND, OR, NOT, XOR) and addition are single-cycle \cite{Lee2020-zo}. BS operates bit-by-bit, leveraging a 1-cycle hardware full adder for arithmetic and achieving zero-cost shifts by simply accessing adjacent rows \cite{Eckert2018-cl}. A critical, costly difference for BS is the lack of a hardware multiplexer (MUX); any conditional logic must be synthesized from four primitive gates, incurring a 4-cycle penalty per bit.

\noindent\textit{Silicon evidence for single-cycle 32-bit addition.}  Our assumption that a 32-bit word-level add completes in one cycle is supported by measured silicon in SRAM compute arrays (e.g., 6T-bitcell datapaths operating at multi-GHz frequencies) and by published BP-IMC designs reporting one-cycle ADD latency~\cite{Lee2020-zo}. We therefore treat word-level logical/arithmetic primitives as single-cycle operations in the BP model.

\begin{table}[h]
\centering
\caption{Compute Cycle Costs for BP and BS Primitives.}
\label{tab:cycle_costs}
\scalebox{1.2}{
\begin{tabular}{l|c||l|c}
\hline
\multicolumn{2}{c||}{\textbf{Bit-Parallel (BP) Mode}} & \multicolumn{2}{c}{\textbf{Bit-Serial (BS) Mode}} \\ \hline
\textbf{Operation} & \textbf{Cycles} & \textbf{Operation} & \textbf{Cycles} \\ \hline
Logic (N-bit) & 1 & 1-bit Add/Sub & 1 \\
ADD (N-bit) & 1 & Shift & 0 \\
SUB (N-bit) & 2 & 1-bit MUX & 4 \\
MULT (N-bit) & N+2 & & \\
SHIFT (k bits) & k & & \\ \hline
\end{tabular}%
}
\end{table}

%Next, we outline \textbf{six key challenges associated with the bit-serial data layout}, each of which contributes to inefficiencies in common computational scenarios.
\subsection{Fundamental Challenges of Bit-Serial Data Layout}

\textit{\textbf{Challenge 1 - Severe Underutilization with Limited Parallelism:}} While PIM is often associated with massively parallel workloads like processing an entire image, many critical, real-time applications involve operating on small, sparse subsets of data. Consider an augmented reality system that must highlight 16 distinct points of interest on a high-resolution video stream. This requires adding a color-correction vector to the 16 corresponding pixel vectors, which is a classic vector addition task. The system must process this small batch of operations with minimal latency to maintain real-time performance, presenting a low degree of parallelism (DoP) workload to the PIM accelerator.

\begin{lstlisting}[language=C, caption={A vector addition kernel with a DoP of 16, analogous to applying color correction to 16 pixels.}, label={lst:low_dop}]
// A[]: original 32-bit pixel color vectors
// B[]: 32-bit color correction vectors
// C[]: resulting pixel color vectors
for (int i = 0; i < 16; ++i) {
    C[i] = A[i] + B[i];
}
\end{lstlisting}

\begin{figure}[h]
\centering
\includegraphics[width=3.4in]{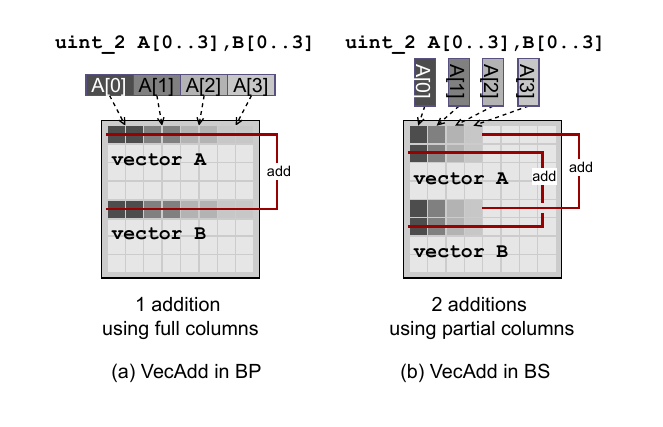}
\caption{Contrasting a 4-element vector addition in (a) Bit-Parallel (BP) and (b) Bit-Serial (BS) layouts. BP configures the array into four wide PEs, utilizing the full array width. BS uses only four 1-bit columns, leaving most of the hardware idle.}
\label{fig:vecadd}
\end{figure}

\textbf{Analysis.} The kernel in Listing~\ref{lst:low_dop} has a DoP of 16. As illustrated conceptually in Figure~\ref{fig:vecadd}b, a BS architecture must dedicate a separate column to each of the 16 parallel operations. This activates only 16 of the 512 available 1-bit PEs, resulting in a dismal $16 / 512 \approx 3.1\%$ resource utilization. The \textbf{BP solution} (Figure~\ref{fig:vecadd}a), however, perfectly matches the hardware to the workload. By configuring the array into 16 PEs, each 32 bits wide for the pixel data, the kernel utilizes all $16 \times 32 = 512$ columns, achieving 100\% resource utilization.

\textit{\textbf{Challenge 2 - Inefficient On-Chip Buffering and Row Overflow:}}
Finite Impulse Response (FIR) filters are a cornerstone of digital signal processing, widely used in applications from audio equalizers to image sharpening and wireless communications. The core computation of an N-tap FIR filter is a convolution: it calculates a weighted sum of the `N` most recent input samples. This inherently requires maintaining a sliding window of past inputs (`x[n]`, `x[n-1]`, etc.) and the filter's coefficients (`b0`, `b1`, etc.) in on-chip memory for fast access. The efficiency of the filtering process thus hinges on using the PIM array as a high-bandwidth, software-managed scratchpad.

\begin{lstlisting}[language=C, caption={Core loop of a 4-tap FIR filter, showing the state update process.}, label={lst:fir}]
// state[]: holds 4 previous input samples (the sliding window)
// coeffs[]: holds 4 filter coefficients
// new_sample: the next input from a data stream

// 1. Update state: shift old samples
state[3] = state[2];
state[2] = state[1];
state[1] = state[0];
state[0] = new_sample;

// 2. Compute output: convolution
output = coeffs[0]*state[0] + coeffs[1]*state[1] + 
         coeffs[2]*state[2] + coeffs[3]*state[3];
\end{lstlisting}

\begin{figure}[h]
\centering
\includegraphics[width=3.4in]{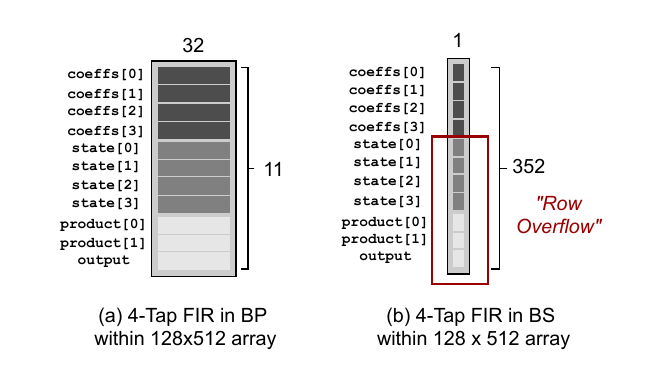}
\caption{Physical data layout for a 4-tap FIR filter. (a) In the BP layout, all required variables—coefficients, state, intermediate products, and the final output—are stored in separate rows, comfortably fitting within the array. (b) The BS layout attempts to store all these variables vertically, causing a massive row overflow.}
\label{fig:fir}
\end{figure}

\textbf{Analysis.} The FIR filter logic in Listing~\ref{lst:fir} requires not only the 4 `state` samples and 4 `coeffs` to be resident, but also space for intermediate products and the final accumulator (at least 11 word-level variables in total, as shown in Figure~\ref{fig:fir}). In a BS layout (Figure~\ref{fig:fir}b), storing these 32-bit words vertically would require $11 \times 32 = 352$ rows. This far exceeds the physical \texttt{array\_rows} of 128, causing a severe \textbf{row overflow} and forcing costly data eviction. The \textbf{BP solution} (Figure~\ref{fig:fir}a) is natural: each 32-bit word occupies a separate row, fitting comfortably within the 128 available rows.

\textit{\textbf{Challenge 3 - Inefficient Intra-Vector Operations:}}
Intra-vector permutations are a critical performance bottleneck in modern cryptography. The Keccak hash function, the winner of the SHA-3 competition, is a prime example. Its $\pi$ stage, a core part of its sponge construction, applies a fixed, non-trivial permutation to the elements of its internal state vector in every round. The efficiency of this data-agnostic shuffling operation is paramount to the overall performance of the algorithm.

\begin{lstlisting}[language=C, caption={Conceptual permutation ($\pi$ stage) in Keccak.}, label={lst:pi_shuffle}]
// state[]: a vector of 25 64-bit words, representing the Keccak state
// next_state[]: a temporary vector to hold the permuted state

// The Pi permutation shuffles elements according to a fixed pattern.
next_state[0] = state[0];
next_state[1] = state[15];
next_state[2] = state[5];
next_state[3] = state[20];
// ... more permutations ...
next_state[22] = state[4];
next_state[23] = state[19];
next_state[24] = state[9];
\end{lstlisting}

\begin{figure}[h]
\centering
\includegraphics[width=3.4in]{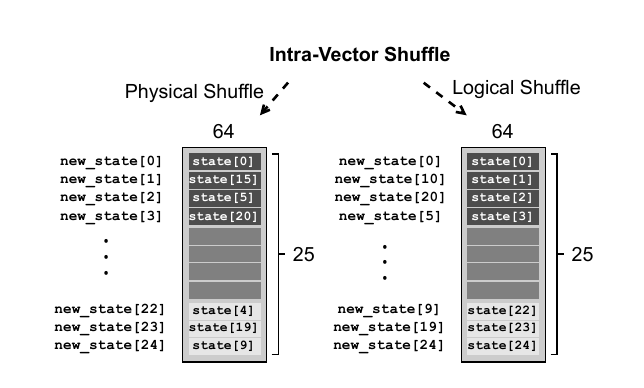}
\caption{Contrasting two permutation mechanisms. Left: a physical shuffle requires explicit data movement between memory locations, incurring multiple read-write cycles. Right: a logical shuffle achieves the same permutation through zero-cost address remapping, native to BP layouts with Element-Serial organization.}
\label{fig:shuffle}
\end{figure}

\textbf{Analysis.} The operational difference is stark, as illustrated in Figure~\ref{fig:shuffle}. The key distinction lies in how permutations are implemented: a \textbf{physical shuffle} requires actual data movement between memory locations, while a \textbf{logical shuffle} achieves the same result through address remapping without moving any data. The left side of Figure~\ref{fig:shuffle} shows a physical shuffle where elements must be explicitly copied from their source locations to new destinations—an operation requiring multiple read-write cycles. The right side demonstrates a logical shuffle where the permutation is achieved by simply updating the mapping between logical indices and physical addresses, incurring zero data movement cost.

In the context of Keccak's $\pi$ permutation, a BS architecture cannot use an Element-Serial (ES) layout due to massive row overflow (25 64-bit elements would require 1,600 rows). It is forced into an Element-Parallel (EP-BS) layout, where elements reside in different columns. Permuting them requires the costly physical shuffle shown on the left—a sequence of explicit inter-column data transfers. In contrast, a BP architecture using an ES-BP layout naturally performs the $\pi$ permutation via the logical shuffle shown on the right, completing the entire operation in zero cycles through address remapping.

\textit{\textbf{Challenge 4 - Inflexible Mixed-Precision Execution:}}
Mixed-precision computation is a cornerstone of efficient deep learning, where models are often quantized to use a mix of 8-bit, 4-bit, and even lower-precision data types to reduce memory and compute costs.

\begin{figure}[h]
\centering
\includegraphics[width=3.4in]{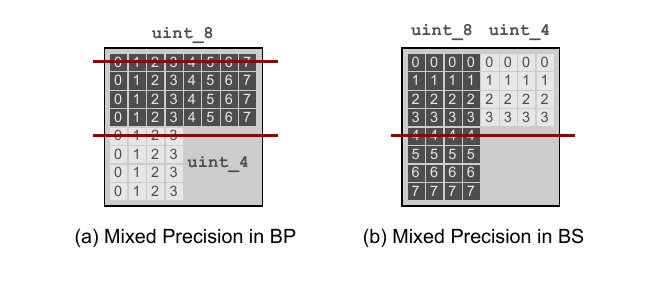}
\caption{Mixed-precision execution. (a) In a BP layout, 8-bit and 4-bit data are processed by independent, word-level PEs. (b) In a BS layout, the bit-level synchronous control creates a conflict: processing bit 4 is valid for the 8-bit data but is out of bounds for the 4-bit data.}
\label{fig:mixed}
\end{figure}

\textbf{Analysis.} The flexibility of the BP layout is shown in Figure~\ref{fig:mixed}(a). Here, 8-bit and 4-bit vectors can be laid out in an ES-BP format. The PIM array is configured into independent, word-level PEs that process their respective data types without conflict. In stark contrast, the BS layout (Figure~\ref{fig:mixed}(b)) faces a fundamental synchronization problem. To process different vectors in parallel, an EP-BS layout is used. When the global controller issues the command to process bit 4, it is a valid operation for the 8-bit data but results in an \textbf{out-of-bitwidth error} for the 4-bit data. This conflict makes true parallel execution of mixed-precision data impractical in BS architectures, forcing inefficient workarounds like padding all data to the largest precision, which negates the benefits of using smaller data types.

\textit{\textbf{Challenge 5 - Impractical Handling of Complex Control Flow:}}
Branching instructions are ubiquitous in general-purpose code. However, PIM architectures, much like GPUs, are ill-suited for traditional control flow where different parallel lanes might take different paths. To avoid breaking parallelism, PIM systems handle branches using \textbf{predicated execution}: both sides of a branch are computed unconditionally, and the correct result is then selected using a flag or a bitmask derived from the branch condition. This transforms a control hazard into a data storage requirement, a trade-off that has profound implications for the underlying data layout.

\begin{lstlisting}[language=C, caption={Predicated execution of a conditional statement.}, label={lst:predicate}]
// Original code: if (A > B) { C = D * E; } else { C = F + G; }

// 1. Compute condition (A > B) and create bitmask
// This is done by checking the sign bit of (B - A).
uint32_t sub = B - A;
// Arithmetic shift right broadcasts the sign bit. If A > B,
// B-A is negative, its sign bit is 1, so the mask becomes all 1s.
uint32_t mask = sub >> 31; // For 32-bit integers

// 2. Compute both branches unconditionally
uint32_t res_true = D * E;
uint32_t res_false = F + G;

// 3. Select the correct result using the mask
C = (res_true & mask) | (res_false & ~mask);
\end{lstlisting}

\textbf{Analysis.} The predicated execution in Listing~\ref{lst:predicate} requires that all operands (A, B, D, E, F, G) and intermediate results (\texttt{sub}, \texttt{mask}, \texttt{res\_true}, \texttt{res\_false}) be stored simultaneously. This amounts to at least 10 word-level variables that must be resident in the PIM array. As demonstrated by the FIR filter example (Challenge 2), this storage requirement is impractical for a BS layout. Storing these 32-bit words vertically would require $10 \times 32 = 320$ rows, causing a severe \textbf{row overflow} of our 128-row array. The \textbf{BP solution}, by contrast, easily accommodates this by storing each variable in a separate row, making predicated execution a viable and efficient strategy.

\textit{\textbf{Challenge 6 - High Inherent Latency for Word-Level Operations:}}
For latency-critical applications such as real-time vehicle control or interactive database queries, the absolute time to complete an operation is more important than aggregate throughput.

\begin{table}[h]
\centering
\caption{Compute Cycle Latency for Common 32-bit Kernels.}
\label{tab:latency_comparison}
\begin{tabular}{l|c|c}
\hline
\textbf{Microkernel} & \textbf{BP Cycles} & \textbf{BS Cycles} \\ \hline
Vector Addition & 1 & 32 \\
Vector Multiplication & 34 & 1024 \\
MIN / MAX & 36 & 192 \\
If-Then-Else & 7 & 97 \\ \hline
\end{tabular}
\end{table}

\textbf{Analysis.} The high latency of BS is not an isolated issue but a fundamental characteristic. As shown in Table~\ref{tab:latency_comparison}, for a 32-bit addition, the BP solution is $32\times$ faster. This latency gap explodes for more complex operations. A 32-bit multiplication in BP takes only 34 cycles, while the bit-serial equivalent requires over 1000 cycles. Even for conditional logic (`If-Then-Else`), BP maintains a significant advantage. For applications where every cycle counts, this dramatic difference in computational latency makes BP the clearly superior architecture.

\subsection{Synthesis: Architectural Root Causes of Bit-Serial's Inefficiencies}
The six challenges detailed previously are not isolated flaws; they are symptoms of four fundamental and intertwined architectural trade-offs inherent to the bit-serial design philosophy:
\begin{itemize}
    \item \textbf{Granularity Mismatch.} The rigid, fine-grained parallelism of the BS architecture is ill-suited for workloads with limited or variable degree of parallelism. This mismatch directly causes the severe resource underutilization demonstrated in \textbf{Challenge 1}, where a vast majority of the hardware remains idle.
    \item \textbf{Vertical Storage Bottleneck.} Storing N-bit words vertically down a column of finite depth is a core tenet of the BS model, but it creates a critical storage bottleneck. This single design choice is the primary source of inefficiency across three distinct scenarios: it renders on-chip buffering for algorithms like FIR filters impractical (\textbf{Challenge 2}); it precludes efficient intra-vector operations such as cryptographic shuffles (\textbf{Challenge 3}); and it makes predicated execution for complex control flow infeasible due to excessive storage demands (\textbf{Challenge 5}). This "row overflow" problem is arguably the most pervasive architectural weakness of the BS paradigm.
    \item \textbf{Lockstep Control Conflict.} The reliance on a single, global control signal broadcast to all 1-bit PEs creates a fundamental conflict when processing heterogeneous data types. This was demonstrated in the mixed-precision case (\textbf{Challenge 4}), where a command valid for one data width is out-of-bounds for another, thus negating the possibility of true parallel execution.
    \item \textbf{Inherent Computational Latency.} By its very definition, bit-serial processing imposes a latency of at least N cycles for any N-bit operation. This results in high latency for all word-level tasks, a weakness starkly quantified by the data in \textbf{Challenge 6}, where the performance gap between BS and BP can be orders of magnitude for common operations.
\end{itemize}

In conclusion, these challenges demonstrate that the implicit "one-size-fits-all" assumption favoring bit-serial layouts is fundamentally flawed. Bit-parallel architectures are not merely an alternative but offer robust solutions to these widespread challenges. The optimal choice of data layout is, therefore, not fixed but is strongly dependent on workload characteristics, demanding a more nuanced, workload-aware approach to PIM system design.

\paragraph{On increasing array rows.}  Some BS limitations (e.g., small working scratchpads) can be alleviated by increasing the number of physical rows. However, this exacerbates array area, wire delay, and sensing energy, and does not address the root causes we identify—vertical storage bottleneck, lockstep control conflict, and inherent bit-serial latency—so the qualitative trade-offs remain.
\section{Evaluation Methodology}\label{sec:methodology}

To provide the first systematic, workload-driven comparison of bit-parallel (BP) and bit-serial (BS) data layouts, we establish a rigorous analytical framework. This framework is grounded in detailed, cycle-accurate performance models of two iso-area Processing-in-Memory (PIM) architectures—one implementing BP execution and the other implementing BS execution—ensuring a fair comparison under identical silicon resource constraints. Our methodology leverages the architectural cost models developed in Section~\ref{sec:arch} to precisely quantify performance across a diverse set of computational patterns, foregoing traditional hardware simulation for a more direct, model-based analysis.

\subsection{Architectural Model}\label{subsec:arch_models}
Our study targets a single 128~row $\times$ 512~column Computing SRAM Array (CSA) that can execute in two mutually exclusive modes: \emph{bit--parallel} (BP) and \emph{bit--serial} (BS).  The physical memory array, sensed by 6T bitcells identical to Lee~\emph{et~al.}~\cite{Lee2020-zo}, is shared; only the 
peripheral logic differs, ensuring an \emph{iso--area} comparison.

\begin{figure}[h]
    \centering
    \includegraphics[width=3.2in]{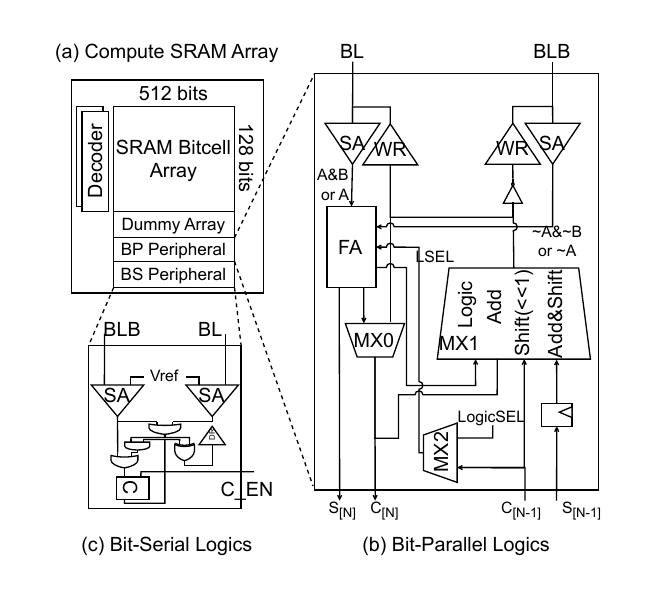}
    \caption{Compute SRAM array \emph{(a)} with dual peripherals: a word--level BP datapath \emph{(b)} and a 1--bit BS datapath \emph{(c)} sharing the same 128~$\times$~512 cell core.  Mode selection is performed by a lightweight mux on each column sense line.}
    \label{fig:csa_dual}
\end{figure}

\subsubsection*{Bit--Parallel (BP) Peripheral}
The BP peripheral aggregates neighbouring bitlines into word--level Processing Elements (PEs).  The array can be \emph{reconfigured at run time} to any word width from 2~to~32~bits, trading off parallelism for precision.  Word--level logic and arithmetic instructions execute in a single cycle; multi--cycle operations such as multiplication follow the cycle costs in Table~\ref{tab:cycle_costs}.  Logical permutations (e.g., Keccak $\pi$) are realised as zero--cycle address remaps.

\subsubsection*{Bit--Serial (BS) Peripheral}
Switching to BS mode disables the word aggregators and exposes each of the 512 columns as an independent 1--bit PE, akin to Neural Cache~\cite{Eckert2018-cl}.  Arithmetic is performed bit by bit with a 1--cycle full adder, and shifts are free by virtue of adjacent rows.  Conditional logic is synthesised from primitives, incurring a 4--cycle \texttt{MUX} penalty per bit.  We model the overhead of intra--vector \emph{physical} shuffles required when data must move across columns.

Because both peripherals attach to the same CSA core, their silicon footprints differ only in peripheral circuits; we conservatively allocate equal metal layers so that total die area remains constant, upholding the iso--area assumption used throughout the evaluation.

\paragraph{Iso--area justification.}
Across SRAM--PIM implementations, published area breakdowns consistently show that the cell array dominates die area (often $>90\%$), while peripheral logic contributes a single--digit percentage. This trend holds for both word--level (BP) and 1--bit (BS) datapaths~\cite{Aga2017-kx,Eckert2018-cl,Lee2020-zo}. Consequently, allocating equal metal and periphery budgets to BP and BS yields a fair, iso--area comparison.

\subsubsection*{On--Chip Transpose Unit}
Many applications benefit from switching between BP and BS representations. We model an on--chip transpose unit attached to the column sense lines that performs bit/word transposition in hardware. The \emph{core} transpose is a single cycle at GHz--class speeds, consistent with prior bitline shuffle hardware~\cite{Hajinazar2021-tu}. End--to--end transpose latency is dominated by array read and write to feed and drain the unit. For an $M$--row logical object in BP and $N$ rows in BS, the total cost is $\text{read}(M) + 1 + \text{write}(N)$ cycles in the BP$\rightarrow$BS direction and $\text{read}(N) + 1 + \text{write}(M)$ in the reverse direction. These costs match the per–round accounting used in our AES study (Section~\ref{sec:tier3}).

\subsection{Model Validation}\label{subsec:model_validation}
Our primitive cycle costs are grounded in measured or reported silicon data from SRAM compute peripheries and in--array logic~\cite{Lee2020-zo,Eckert2018-cl,Hajinazar2021-tu}. We validate our analytical totals against two public benchmark suites, MIMDRAM~\cite{Oliveira2024-fr} (micro–kernels) and PIMBench~\cite{Siddique2024-ip} (applications). The relative trends we report agree with the layout–specific results produced by the PIMeval framework~\cite{Siddique2024-ip}, providing confidence in the model's fidelity.

\subsection{Benchmark Suite}\label{subsec:benchmarks}
To comprehensively evaluate the trade-offs between BP and BS, we employ a curated two-tier benchmark suite designed to expose the layout affinities of diverse computational patterns.

\subsubsection{Tier 1: Microbenchmarks}
To isolate the performance of fundamental computational patterns, we select a broad set of microbenchmarks inspired by the MIMDRAM~\cite{Oliveira2024-fr} suite. These kernels include vector arithmetic, logical and comparison operations, control flow primitives, and data organization patterns like reductions. These microbenchmarks allow us to directly measure and validate the performance characteristics and latency costs detailed in our architectural models.

\subsubsection{Tier 2: Application Benchmarks}
To assess performance on real-world workloads, we draw from the PIMBench~\cite{Siddique2024-ip} suite, which covers a wide range of application domains. Our selection includes representative kernels from Deep Learning (VGG), Cryptography (Keccak), Signal Processing (FIR filters), and Data Analytics (database-style queries). By analyzing these full applications, we can demonstrate how the performance of individual computational patterns aggregates at the workload level, thereby providing the evidence needed to develop our final layout selection guidelines.

\section{Evaluation}
\label{sec:eval}

\subsection{Evaluation Goals}
\label{sec:eval-goals}

Section~\ref{sec:arch} distilled \emph{six concrete challenges} of the canonical bit-serial (BS) layout and traced them back to \emph{four architectural root causes}: (\textit{i}) granularity mismatch, (\textit{ii}) vertical storage bottleneck, (\textit{iii}) lock-step control conflict, and (\textit{iv}) inherent computational latency.  
Our evaluation therefore aims to determine \emph{when and why} each data layout succeeds or fails with respect to these root causes. We use a two-tier benchmark suite that couples micro-kernels and full applications.

We address the following three questions:

\begin{enumerate}
    \item \textbf{Q1 — Layout \(\rightarrow\) Kernel Mapping:}\\
    For the diverse micro-kernels that dominate modern workloads—arithmetic, logical, reduction, and control-intensive operations—how do BP and BS compare in compute latency, and which root cause is the primary limiter for each kernel category?
    
    \item \textbf{Q2 — Density vs.\ Latency Trade-off:}\\
    As data sets grow beyond a single array's capacity, batching becomes mandatory. How does the tension between BP's lower storage density and BS's higher per-operation latency influence end-to-end execution time, and where is the inflection point where batching overhead neutralises BP's compute advantage?
    
    \item \textbf{Q3 — Application-Level Impact and Adaptivity:}\\
    When kernels with conflicting layout preferences coexist within one application (e.g., AES combining table look-ups and matrix multiplications), what is the net performance of a \emph{static} BP or BS choice? Can a \emph{hybrid} strategy that switches layouts at phase boundaries offset individual weaknesses despite transposition overheads?
\end{enumerate}

Answering these questions provides workload-aware guidelines for selecting—or dynamically combining—data layouts, validating our claim that no single layout is universally optimal across the PIM design space.

\subsection{Experimental Setup}
\label{sec:eval-setup}

\textbf{Hardware Assumptions.}  
All results are obtained with the iso-area bit-parallel (BP) and bit-serial (BS) array organizations defined in Section~\ref{sec:methodology}.  
Unless stated otherwise, we use the baseline configuration of 128~rows and 512~columns per array (Table~\ref{tab:arch_params}) and the cycle-level cost model of Table~\ref{tab:cycle_costs}.  
Both layouts share the same total silicon footprint; differences in performance stem solely from data placement and execution style.

\textbf{Benchmark Suite.}  
We directly reuse the two-tier benchmark suite introduced in Section~\ref{subsec:benchmarks}.  
Tier~1 comprises 18 micro‐kernels that exercise arithmetic, logical, reduction, and control‐flow patterns, evaluated with two input sizes: \emph{small} (1\,024 elements) and \emph{large} (65\,536 elements).  
Tier~2 contains 22 full applications spanning vision, learning, analytics, and cryptography; each uses widely adopted dataset dimensions (e.g., CIFAR-10 for VGG‐16, 1 M points for K-means) and identical inputs across layouts.

\textbf{Methodology.}  
For each layout we compute total latency as  
\(\text{cycles} = \text{load} + \text{compute} + \text{readout}\),  
where \textit{compute} cycles come directly from the primitives in Table~\ref{tab:cycle_costs}.  
Load and readout costs scale linearly with the number of active rows and therefore reflect batching overhead when BP overflows the array.  
End-to-end runtimes are normalised to a 1\,GHz array clock and reported in cycles for maximum portability.

\subsection{Tier--1: Microbenchmark Analysis}
\label{sec:eval_micro}

Our microbenchmark analysis proceeds in two steps, using precise data from Tables~\ref{tab:micro_detailed} and~\ref{tab:latency_density}. First, we examine performance at a fixed scale (1024 elements) to understand core computational trade-offs. Second, we analyze sensitivity to workload size to reveal the critical impact of data layout density and batching.

\paragraph{Performance at Fixed Scale.}
Table~\ref{tab:micro_detailed} provides a cycle-level breakdown that reveals a wide performance spectrum, directly tracing back to the architectural challenges.

The \emph{Arithmetic} cluster highlights the raw efficiency of BP's word-level datapath. For basic operations like \texttt{Vector Add}, BP's `Compute` time is just 1-2 cycles, while BS requires 16, a direct result of its \textbf{Inherent Latency (Challenge 6)}. The gap widens for \texttt{MULTU}, where BP's 18-cycle hardware is over 14$\times$ faster than the 256-cycle BS shift-and-add implementation.

The \emph{Logical / Bit-manipulation} cluster reveals a more nuanced picture. For bit-centric tasks like \texttt{bitcount}, the \textbf{Granularity Mismatch (Challenge 1)} makes BP less efficient; its total latency (185 cycles) is higher than the more natural serial summation in BS (128 cycles). Conversely, the native serial `Reduction` in BS is highly optimized and slightly outperforms the BP tree-based version.

The \emph{Control / Predicate} cluster demonstrates BP's performance advantage for conditional operations. Kernels such as \texttt{abs} and \texttt{if-then-else} require conditional logic that BS must synthesize through multiplexers, increasing compute cycles from 7 (BP) to 49 (BS) for \texttt{if-then-else}—a consequence of the \textbf{Lock-step Control Conflict (Challenge 4)}. These kernels also exhibit the \textbf{Vertical Storage Bottleneck (Challenge 2)}: the \texttt{if-then-else} kernel's 10 live variables require 52 rows in BS versus 5 in BP, indicating potential overflow issues in complex predicated code.

\paragraph{Sensitivity to Workload Size.}
The fixed-scale analysis does not capture the full picture. End-to-end performance is critically sensitive to workload size due to BP's lower storage density. Table~\ref{tab:latency_density} uses vector addition to illustrate this "batching effect" with precise figures.

\begin{table}[htp]
    \centering
    \caption{Vector addition latency as a function of workload size.  BP batches increase once the working set exceeds 16K elements, neutralising its compute advantage.  Speedup is $\text{BS}/\text{BP}$.}
    \label{tab:latency_density}
    \scalebox{0.9}{
    \begin{tabular}{r|c|c|c|c}
        \hline
        \textbf{Elements} & \textbf{BP Batches} & \textbf{BP Cycles} & \textbf{BS Cycles} & \textbf{Speedup} \\
        \hline
        1K     & 1  & 97    & 112   & 1.15$\times$ \\
        4K     & 1  & 385   & 400   & 1.04$\times$ \\
        16K    & 1  & 1,537 & 1,552 & 1.01$\times$ \\
        64K    & 4  & 6,148 & 6,160 & 1.00$\times$ \\
        256K   & 16 & 24,592 & 24,592 & 1.00$\times$ \\
        \hline
    \end{tabular}}
\end{table}

For 1K-element workloads, BP achieves a 1.15$\times$ speedup over BS. This advantage decreases to 1.01$\times$ at 16K elements—BP's single-batch capacity—as data movement costs (load/readout) become significant. At 64K elements, BP requires 4 sequential batches, and the overhead of repeated data loading eliminates its computational advantage, resulting in performance parity with single-batch BS execution. This demonstrates the fundamental trade-off between BP's low latency (Challenge 6) and its lower storage density (Challenge 2).

\paragraph{Synthesis.}
The results demonstrate that optimal layout selection depends on both \emph{kernel characteristics} and \emph{workload size}. At small scales, kernel-specific properties determine performance: BP achieves up to 14$\times$ speedup for arithmetic operations (\texttt{MULTU}), while BS performs 1.4$\times$ better for bit-level operations (\texttt{bitcount}). As workload size increases, BP's lower storage density necessitates batching, which diminishes or eliminates its computational advantage. These findings support workload-aware PIM architectures over static layout choices.

% --------------------------------------------------------------------------
% Detailed latency breakdown table added for completeness
% --------------------------------------------------------------------------

\begin{table*}[t]
    \centering
    \caption{Cycle breakdown of advanced micro‐kernels.  Rows/Elem and Cols/Elem denote physical footprint per element in the array.  All values are for 32-bit data unless noted.}
    \label{tab:micro_detailed}
    \footnotesize
    \setlength{\tabcolsep}{5pt}
    \resizebox{\textwidth}{!}{%
    \begin{tabular}{l|llccccc|c|l}
        \hline
        \textbf{Kernel} & \textbf{Variant} & \textbf{Mode} & \textbf{Rows/Elem} & \textbf{Cols/Elem} & \textbf{Load} & \textbf{Compute} & \textbf{Readout} & \textbf{Total} & \textbf{Challenge} \\
        \hline
        \hline
        \multicolumn{10}{c}{\emph{Arithmetic kernels}} \\
        \hline
        Vector~Add & Standard & BP & \textasciitilde3 & 16 & 64 & 1 & 32 & 97 & 6 \\
                    &         & BS & 49 & 1 & 64 & 16 & 32 & 112 & 6 \\
        \hline
        Vector~Sub & Standard & BP & \textasciitilde3 & 16 & 64 & 2 & 32 & 98 & 6 \\
                    &         & BS & 49 & 1 & 64 & 16 & 32 & 112 & 6 \\
        \hline
        MULTU & HW & BP & \textasciitilde4 & 16 & 128 & 18 & 64 & 210 & 6 \\
                & Shift+Add & BS & 64 & 1 & 64 & 256 & 64 & 384 & 6 \\
        \hline
        MULTU~Const & HW & BP & \textasciitilde3 & 16 & 128 & 18 & 64 & 210 & 6 \\
                     & Shift+Add & BS & 48 & 1 & 64 & 256 & 64 & 384 & 6 \\
        \hline
        DIVU & Restoring & BP & 4 & 16 & 64 & 640 & 32 & 736 & 6 \\
              & Restoring & BS & 64 & 1 & 64 & 1280 & 32 & 1376 & 6 \\
        \hline
        MIN & Shift Mask & BP & \textasciitilde5 & 16 & 64 & 21 & 32 & 117 & 6 \\
            & Iter. Comp. & BS & 50 & 1 & 64 & 96 & 32 & 192 & 6 \\
        \hline
        MAX & Shift Mask & BP & \textasciitilde5 & 16 & 64 & 21 & 32 & 117 & 6 \\
            & Iter. Comp. & BS & 50 & 1 & 64 & 96 & 32 & 192 & 6 \\
        \hline
        \multicolumn{10}{c}{\emph{Logical / Bit--manipulation kernels}} \\
        \hline
        Reduction & Tree & BP & 2 & 16 & 32 & 19 & 16 & 67 & 6 \\
                   & Native & BS & 17 & 1 & 32 & 16 & 16 & 64 & 6 \\
        \hline
        bitcount & D\&C & BP & 3 & 16 & 128 & 25 & 32 & 185 & 1 \\
                  & Summation & BS & 26 & 1 & 32 & 80 & 16 & 128 & 1 \\
        \hline
        bitweave & 1b Logic & BP & 53 & 1024 & 96 & 225 & 2 & 323 & 1 \\
                  & 2b Logic & BS & 74 & 512 & 64 & 434 & 2 & 500 & 1 \\
                  & 4b Logic & BS & 116 & 256 & 48 & 852 & 2 & 902 & 1 \\
        \hline
        \multicolumn{10}{c}{\emph{Control / Predicate kernels}} \\
        \hline
        abs & Shift Mask & BP & 3 & 16 & 32 & 18 & 32 & 82 & 4 \\
            & Serialised & BS & 48 & 1 & 32 & 48 & 32 & 112 & 4 \\
        \hline
        if\texttt{-}then\texttt{-}else & Mask 0-s & BP & 5 & 16 & 96 & 7 & 32 & 135 & 2/6 \\
                                        & Synth.~MUX & BS & 52 & 1 & 80 & 49 & 32 & 161 & 2/6 \\
        \hline
        equal & XOR+Reduce & BP & 3 & 16 & 64 & 22 & 32 & 118 & 6 \\
               & Serial XOR & BS & 49 & 1 & 64 & 33 & 32 & 129 & 6 \\
        \hline
        ge\texttt{\_}0 & Shift & BP & 1 & 16 & 32 & 17 & 16 & 65 & 6 \\
                 & Sign Bit & BS & 16 & 1 & 32 & 1 & 16 & 49 & 6 \\
        \hline
        gt\texttt{\_}0 & Synth. & BP & 3 & 16 & 32 & 35 & 32 & 99 & 6 \\
                 & Serial Red. & BS & 17 & 1 & 32 & 17 & 16 & 81 & 6 \\
        \hline
        ReLU (8K) & Standard & BP & \textasciitilde2 & 16 & 512 & 17 & 512 & 1041 & 4 \\
                   & Standard & BS & \textasciitilde17 & 1 & 512 & 17 & 512 & 1041 & 4 \\
        \hline
    \end{tabular}}
\end{table*}

\paragraph{Implication—batching neutralizes advantages.}
The measurements confirm that layout performance depends on both kernel characteristics and workload size. BP's latency advantage applies only when the working set fits within a single array batch. When batching is required, data movement costs dominate execution time, causing both layouts to achieve similar performance. This size sensitivity explains the reduced speedups observed in application-level benchmarks (Section~\ref{sec:tier3}) compared to micro-kernels, reinforcing the need for workload-aware PIM architectures.

\subsection{Tier--2: Application Benchmark Results}
\label{sec:tier3}

Micro‐kernels provide architectural intuition, but real programs combine many
kernel types and exhibit diverse working‐set footprints.  We therefore profile
22 end‐to‐end applications drawn from vision, learning, analytics, and
cryptography.  The performance metric is total kernel cycles under the BP and
BS layouts, using the analytical framework of
Section~\ref{sec:eval-setup}.  To expose the abundant thread- and data-level
parallelism of these real‐world workloads, we assume a system with \textbf{512
parallel arrays}, scaling the core architecture of
Section~\ref{sec:methodology} accordingly.
Table~\ref{tab:app_class} groups the workloads
by their observed speedup and dominant limiting factor.

\begin{table*}[t]
    \centering
    \caption{Classification of application benchmarks.  Speedup is reported as
    $\text{BS}/\text{BP}$; values $<1$ indicate BS is faster.}
    \label{tab:app_class}
    \scalebox{0.95}{
    \begin{tabular}{l|l|c|l}
        \hline
        \textbf{Category} & \textbf{Applications} & \textbf{Speedup (BS/BP)} & \textbf{Dominant Architectural Factor} \\
        \hline
        Strong BP preference & Brightness, K-means, Keccak, FIR & 1.5--3.0$\times$ & Mixed arithmetic / control (Challenges 4,6) \\
        Moderate BP preference & VGG-13, VGG-16/19, GEMM, GEMV, Conv, Downsample & 1.2--1.5$\times$ & High arithmetic intensity, limited batching (6) \\
        Balanced & Vector-Add, AXPY, Pooling, Prefix-Sum & 1.0--1.15$\times$ & Batching neutralises latency (2) \\
        BS preference & Histogram, HDC, Bitweave-DB & 0.6--0.9$\times$ & Bit-centric, full-density layouts (1) \\
        Hybrid recommended & AES, Radix-Sort & 2.66$\times$ speedup$^*$ & Phase diversity (3,4,5) \\
        \hline
    \end{tabular}}
    % \vspace{-0.5em}
    {\footnotesize $^*$Hybrid speedup is relative to the best static layout (BP for AES).}
\end{table*}

\begin{figure}[h]
    \centering
    \includegraphics[width=0.9\columnwidth]{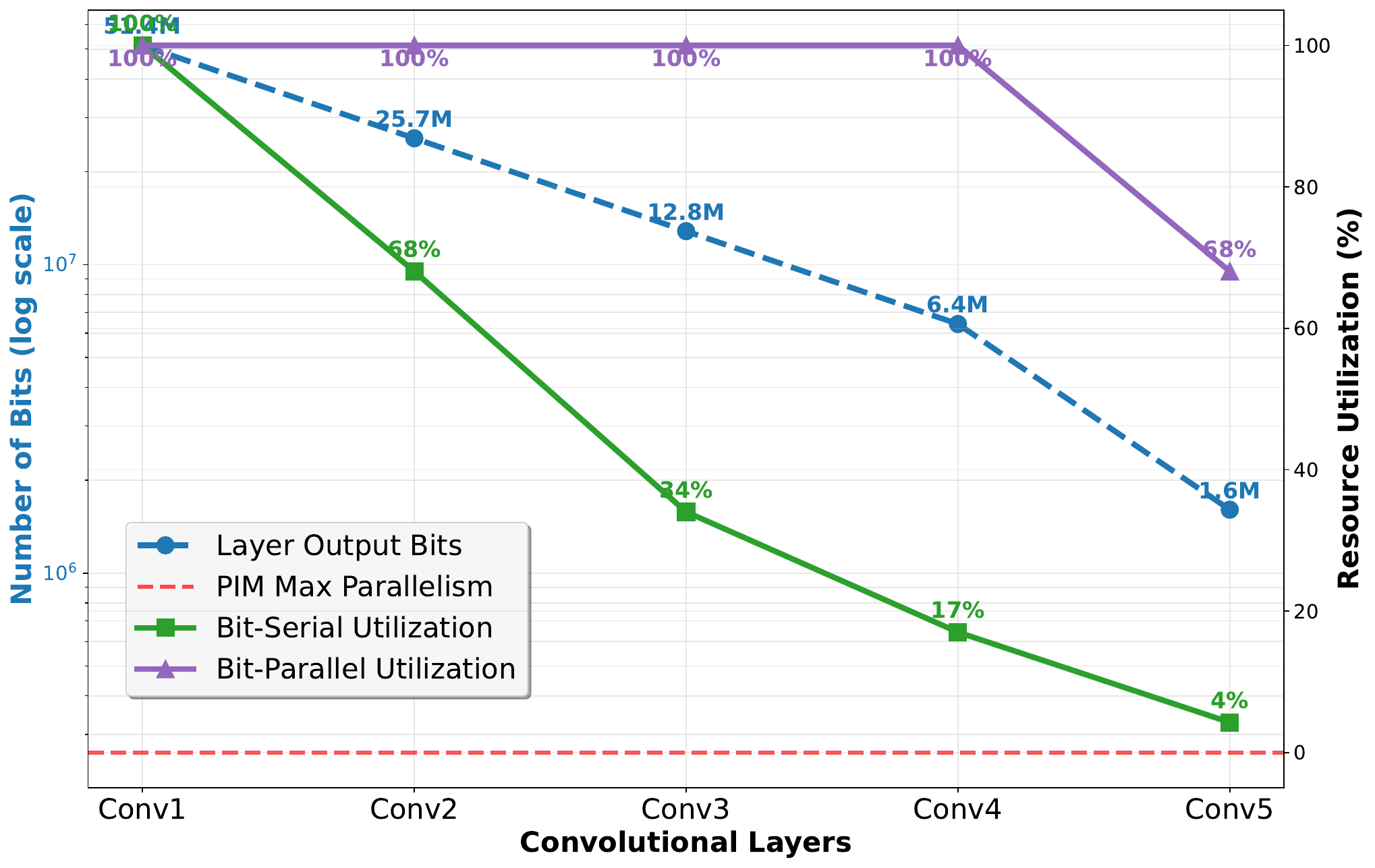}
    \caption{VGG-13 layer output size and PIM resource utilization. Left axis (log scale) shows layer output bits versus PIM capacity. Right axis shows the resulting resource utilization for BP and BS.}
    \label{fig:vgg13_util}
\end{figure}

\paragraph{Case study~1: VGG-13 inference (resource utilization).}
VGG-13 inference \cite{Simonyan2014-my} demonstrates how resource utilization varies across convolutional layers as feature map dimensions decrease. Figure~\ref{fig:vgg13_util} quantifies this effect by comparing each layer's output size (assuming 16-bit elements and 3$\times$3 kernel reuse) against the PIM's maximum parallelism of 262,144 bits.

The utilization patterns reveal fundamental architectural differences:
\begin{itemize}
    \item \textbf{Early layers (Conv1-Conv3):} Both layouts achieve 100\% utilization, as the feature maps exceed the PIM's processing capacity. The performance difference stems solely from BP's lower computational latency (Challenge 6).
    
    \item \textbf{Later layers (Conv4-Conv5):} BS utilization decreases to 17\% (Conv4) and 4\% (Conv5) due to insufficient parallelism—the 16-bit element count cannot saturate 262,144 bit-serial units. BP maintains higher utilization (100\% for Conv4, 68\% for Conv5) through its word-level granularity, which better matches the reduced workload dimensions.
\end{itemize}

These measurements demonstrate that BS suffers from the \textbf{Granularity Mismatch (Challenge 1)} in later network layers. While early layers provide sufficient parallelism for both layouts, the progressive reduction in feature map sizes leaves BS compute resources increasingly underutilized. This quantitative evidence contradicts the assumption that BS universally provides superior efficiency in neural network inference.

\paragraph{Case study~2: AES-128 encryption (hybrid wins).}
AES-128 encryption combines four distinct operations with strongly conflicting layout preferences, making it an ideal showcase for hybrid execution. Table~\ref{tab:aes_perf} quantifies the dramatic performance disparities across AES stages.

\begin{table}[h]
    \centering
    \caption{Per-round cycle breakdown for AES-128 operations.}
    \label{tab:aes_perf}
    \scalebox{0.85}{
    \begin{tabular}{l|c|c|c}
        \hline
        \textbf{Operation} & \textbf{Bit-Parallel} & \textbf{Bit-Serial} & \textbf{Best Layout} \\
        \hline
        AddRoundKey & 16 & 128 & BP (8$\times$) \\
        SubBytes & 1,568 & 115 & BS (13.6$\times$) \\
        ShiftRows & 32 & 256 & BP (8$\times$) \\
        MixColumns & 272 & 2,176 & BP (8$\times$) \\
        \hline
        \textbf{Total per round} & 1,888 & 2,675 & - \\
        \hline
    \end{tabular}}
\end{table}

The performance disparity stems from fundamental architectural mismatches:
\begin{itemize}
    \item \textbf{SubBytes:} The S-box lookup is the most expensive operation in BP, requiring complex GF($2^8$) inversion implemented via composite field arithmetic. This consumes approximately 98 cycles per byte, totaling 1,568 cycles for 16 bytes. BS, however, leverages the Boyar-optimized bit-sliced implementation \cite{Boyar2010-xe} that transforms the S-box into just 115 logic gates, completing in 115 cycles—a remarkable 13.6$\times$ speedup.
    \item \textbf{Other stages:} AddRoundKey (XOR), ShiftRows (byte permutation), and MixColumns (GF multiplication) are naturally word-oriented operations where BP excels with its native 8-bit datapath, achieving 8$\times$ better performance than BS.
\end{itemize}

A \emph{hybrid} execution strategy capitalizes on these complementary strengths: execute SubBytes in BS, all other operations in BP, with layout transpositions between phases. The total cost becomes:
\[ \text{Per round: } 16 + 115 + 32 + 272 + 290 = 725 \text{ cycles} \]
where the 290-cycle transposition overhead\footnote{The 290-cycle cost includes two transpositions: BP$\rightarrow$BS (145 cycles) before SubBytes and BS$\rightarrow$BP (145 cycles) after. Each 145-cycle transposition breaks down as follows: (1) reading data to the transposer: 16 cycles for BP$\rightarrow$BS or 128 cycles for BS$\rightarrow$BP, (2) hardware transpose operation: 1 cycle, and (3) writing back: 128 cycles for BP$\rightarrow$BS or 16 cycles for BS$\rightarrow$BP. The asymmetry reflects the different row counts: AES state occupies 16 rows in BP (1 bytes/row × 16 rows) but 128 rows in BS (1 bit/row × 128 bits).} is amortized across the round. 

\textbf{Sensitivity to transpose cost.}  Our conclusions are robust to slower transpose hardware. If we conservatively increase the transpose cost by $10\times$ (core latency from 1 to 10 cycles), the total AES runtime increases by only $\sim$2.6\%, and the hybrid schedule still achieves a 2.59$\times$ speedup over the best static layout (BP). Thus, the hybrid benefit is insensitive to transpose latency within a wide, practical range.

For the complete AES-128 encryption (10 rounds):
\begin{itemize}
    \item Pure BP: 18,624 cycles (SubBytes dominates at 83.1\% of runtime)
    \item Pure BS: 26,750 cycles (MixColumns becomes the bottleneck)
    \item Hybrid: 6,994 cycles (balanced execution with transposition overhead)
\end{itemize}

This \textbf{2.66$\times$} improvement over the best static layout (BP) demonstrates that judicious layout switching can overcome the individual weaknesses of each paradigm. The win is particularly compelling for AES because: (1) SubBytes has an extreme 13.6$\times$ performance advantage in BS due to bit-sliced optimization, (2) the transpose cost (2,900 cycles total) is only 20\% of the compute savings (14,530 cycles), and (3) the regular round structure enables predictable, compiler-driven layout management.

\paragraph{Takeaway.}  Application results corroborate the kernel analysis:
(1) BP's latency advantage translates to\,\(\sim1.3\times\) median speedup for
arithmetic-heavy workloads; (2) bit-centric tasks favour BS despite its higher
per-operation latency; (3) hybrid execution can exceed either static layout
when phase diversity is high.  These findings motivate future PIM designs that
support fast, low-energy layout transposition rather than locking the system to
one paradigm.

\paragraph{Energy considerations.}  While we defer a full energy model to extended work, measured silicon data already suggests that word-parallel SRAM-PIM can be highly energy-efficient. Reported TOPS/W for ADD in SRAM PIM shows BP-style datapaths achieving higher energy efficiency than BS-style ones in comparable technologies (e.g., $\sim$8.1~TOPS/W \cite{Lee2020-zo} vs.~$\sim$5.3~TOPS/W \cite{Wang2020-wk} for ADD). Combined with our latency analysis, these figures indicate that the most energy-efficient layout is workload-dependent, and hybrid strategies that minimise time spent in an energy-inefficient layout can further reduce energy.

\subsection{Synthesis and Key Takeaways}
\label{sec:eval-synth}

The three evaluation tiers paint a coherent picture: neither data layout
unilaterally dominates; instead, each excels under specific workload
conditions.  Table~\ref{tab:recap} summarises the qualitative trends distilled
from more than forty quantitative data points.

\begin{table}[htp]
    \centering
    \caption{Workload characteristics favoring each layout.}
    \label{tab:recap}
    \scalebox{0.88}{
    \begin{tabular}{p{0.45\columnwidth}|p{0.45\columnwidth}}
        \hline
        \textbf{BP-friendly} & \textbf{BS-friendly} \\
        \hline
        Word-level arithmetic (add/mul/div) & Bit-level operations (popcount, XOR) \\
        Conditional logic, predication & Uniform, data-independent control \\
        Mixed precision vectors & High DoP, full utilization \\
        Latency-critical tasks & Large working sets \\
        Low degrees of parallelism & Logical transpositions \\
        \hline
    \end{tabular}}
\end{table}

\paragraph{Hybrid data layouts as the next frontier.}  The AES case study
underscores that dynamic layout switching---enabled by an on--chip transpose
engine---can beat the best static choice.  Our analytical model shows that when
transpose cost is below~\(2\%\) of per--phase runtime (51~cycles in our
configuration), a hybrid schedule is profitable for any application that
contains at least one BS--favourable and one BP--favourable phase.  This opens
an attractive design space between the two historical extremes.

\paragraph{Future work.}  Two directions merit further investigation: (1)~fine
-grained, row--selective transpose units that amortise cost over partial data
sets; and (2)~compiler analyses that automatically partition code into
layout--optimal regions and insert transpose operations only when the expected
gain exceeds the hardware threshold.  We leave the exploration of these
co--optimised hardware–software techniques to future work.

In summary, our evaluation confirms the central thesis: 
\emph{data layout is a first--order design decision for PIM}.  A workload--aware
architecture that flexibly navigates the BP/BS continuum stands to unlock
significant, otherwise untapped performance.  The methodology and empirical
findings presented here provide a quantitative foundation for such adaptive
systems.

\section{Conclusion}
\label{sec:conclusion}

This paper presented the first systematic comparison of bit-serial (BS) and bit-parallel (BP) data layouts in Processing-in-Memory architectures. Through cycle-accurate modeling and evaluation across 22 applications, we identified six fundamental challenges of BS layouts and delivered three key insights:
\textbf{Architecture matters.} BP achieves up to 14$\times$ lower latency than BS for arithmetic and control-heavy kernels by eliminating serial carry propagation.
\textbf{Density matters more.} When workloads exceed array capacity, BP's batching overhead neutralizes its computational advantages, enabling BS to excel on bit-centric or large-scale problems.
\textbf{Flexibility wins.} Hybrid execution that switches layouts at phase boundaries outperforms static choices by up to 2.66$\times$ when transposition cost is minimal.
These findings transform data layout from an implicit assumption into a first-class design parameter. Future PIM systems should support low-cost transposition hardware and workload-aware compilers that orchestrate dynamic layout decisions. By establishing that neither BP nor BS is universally optimal, this work motivates adaptive architectures that navigate the BP/BS continuum based on workload characteristics.

\bibliographystyle{IEEEtranS}
\bibliography{paperpile,references-ES}

\end{document}